# Formulation of the partition functions and magnetizations for two-dimensional nearest neighbour Ising models for finite and infinite lattice sites

Anshu Priya and M.V. Sangaranarayanan*

Department of Chemistry

Indian Institute of Technology -Madras

Chennai 600036 India

sci.anshu@gmail.com and sangara@iitm.ac.in

## ABSTRACT

Using a combinatorial method, the partition functions for two-dimensional nearest neighbour Ising models have been derived for a square lattice of 16 sites in the presence of the magnetic field. A novel hierarchical method of enumeration of all the configurations for any arrangement of sites has been proposed. This enumeration has been executed by a systematic analysis of the appropriate diagrams without employing any algorithmic approach or computational tools. The resulting algebraic eqn in terms of the magnetic field and nearest neighbour interaction energies may then provide a methodology for deducing the magnetization in the thermodynamic limit of infinite sites. A semi-empirical eqn for magnetization is proposed for non-zero magnetic fields.

## 1.Introduction

The formulation of the spontaneous magnetization pertaining to two-dimensional nearest neighbour Ising models constitutes a significant milestone in condensed matter physics [1]. Since the one-dimensional nearest neighbour Ising models do not predict phase transitions, the two-dimensional analogues have subsequently evoked considerable interest [2-4]. Onsager's exact solution of two-dimensional Ising models is a break-through in this context which yielded an explicit expression for estimating the critical temperatures [5]. Further insights regarding the Onsager's exact solution resulted from the analysis of Yang [6]. Among various approximations investigated in this context, mention should be made of the following: Bragg-Williams Approximation (BWA) [7], Bethe quasi-chemical approach [8], series expansions [9], renormalization group techniques [10], scaling hypothesis [11], graph theoretical procedures [12] etc. The results derived in the context of Ising models become applicable *mutatis mutandis* to various topics in solid state and condensed matter physics on account of the equivalence with binary alloys and lattice gas description of fluids [13]. In particular, order-disorder transitions [14], condensation at electrochemical interfaces [15], phase separation in self-assembled monolayer films [16], protein folding [17], free energies of surface steps [18] etc. deserve mention among several applications of the two-dimensional Ising models [19]. Hence Ising model continues to be the most celebrated model [20] and its exact solution is considered as Holy Grail.

While rigorous methodologies are always essential for grasping the complexity of Ising models, any heuristic analysis is eminently desirable, since it provides fascinating insights-not transparent within the realm of mathematical rigor. In this Communication, new equations for the zero-field partition function and spontaneous magnetization are derived for two-dimensional square lattice Ising models. The objectives of this Communication are to (i) provide analytical formula for the enumeration of all types of black-white edges for a square lattice of 16 sites;(ii) deduce the canonical partition function by employing the methodology and (iii) postulate an explicit eqn for field-dependent magnetization in a semi-empirical manner.

## 2. Partition function for a square lattice of 16 sites

Consider the two-dimensional nearest neighbour Ising model on a square lattice [4] where the Hamiltonian is represented as

$$H_T = -J\sum_{<ij>}(\sigma_i\,\sigma_{i+1} + \sigma_i\sigma_{i+1,j}) - H\sum\sigma_i \qquad (1)$$

where $J$ denotes the nearest neighbour interaction energy, $H$ being the external magnetic field, $i$ and $j$ denote the row and column index, respectively. The spin variable $\sigma_{i,}$ assumes the value of+1 or -1 in the spin ½ Ising model. The corresponding canonical partition function is given by [21]

$$Q = \sum_{\sigma_i} \sum_{\sigma_{i+1}} \dots \sum_{\sigma_N} e^{\frac{-J}{kT}\{\sigma_N\}} \tag{2}$$

where $k$ denotes Boltzmann constant, $T$ being the absolute temperature. Several simulation strategies exist for large lattices which yield numerical estimates for the partition function, internal energy, specific heat and magnetization [22].

The mapping of the Ising model problem to the concepts of the graph theory is well known [23] .The basic tenet in this context is the visualization of the nearest neighbour Ising model as a bipartite graph G = $\{V, E\}$consisting of the vertices $V$ and edges $E$ .The vertices represent the arrangement of +1 (designated as 'black') and -1(designated as 'white') spins, while the edges correspond to the connectivity between the vertices and may be correlated to the interaction energies. For brevity, we assume that the symbol '$p$' denotes the number of +1 spins (black sites) and *'q'* represents the number of black-white edges. $A(p, q)$ denotes the number of times a configuration occurs with '$p$' black sites and *'q'* black-white (B-W)edges. Thus, the estimation of the partition function involves essentially the counting of the B-W edges; however, it should be noted that even for 16 sites, this enumeration entails the listing of 65536 states; this is not all; for obtaining the critical properties, the thermodynamic limit is a *sine qua non* , warranting these computations of B-W edges for infinite sites.

## 2.1 Enumeration of the black-white edges

Our methodology of enumeration involves progressive incorporation of additional sites to earlier arrangements in a systematic manner. The detailed steps are indicated for the calculation of A(p,q) when p =8 and q =16 *( vide supra)*

Figure 1 provides a detailed description of the arrangements for various random distribution of sites. It is a non-trivial endeavour to formulate a closed form expression that yields the correct entries for each value of p and q.

**Figure 1: Schematic representation of typical arrangements for progressive incorporation of the sites and the corresponding edges**

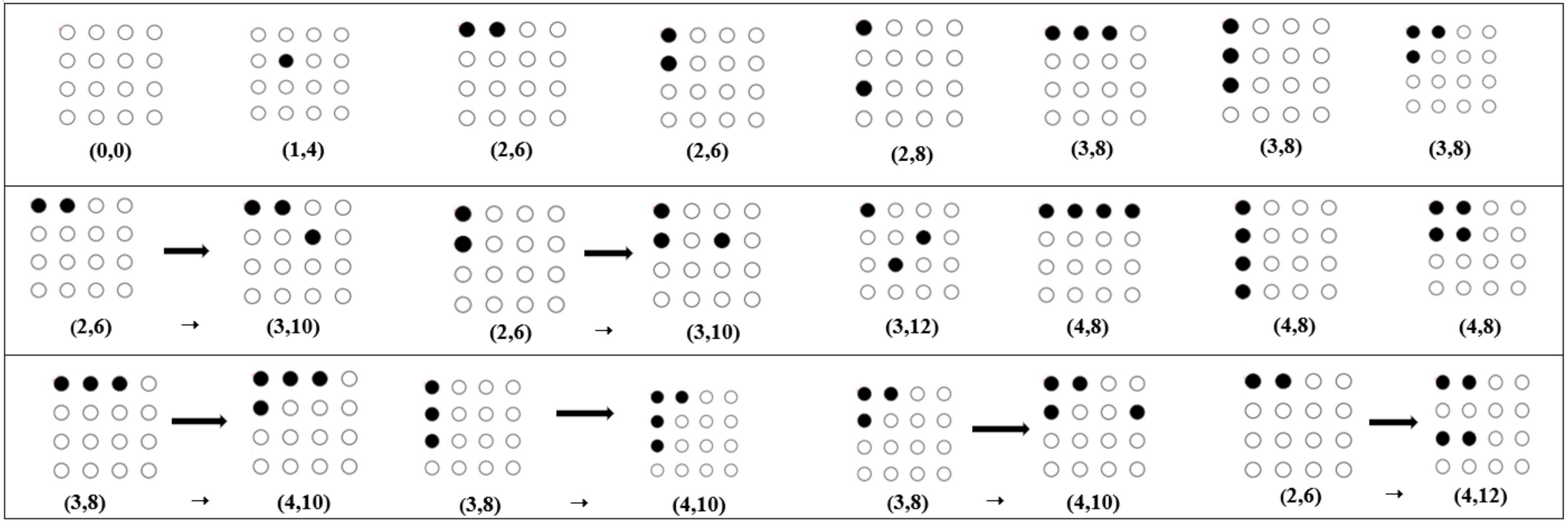

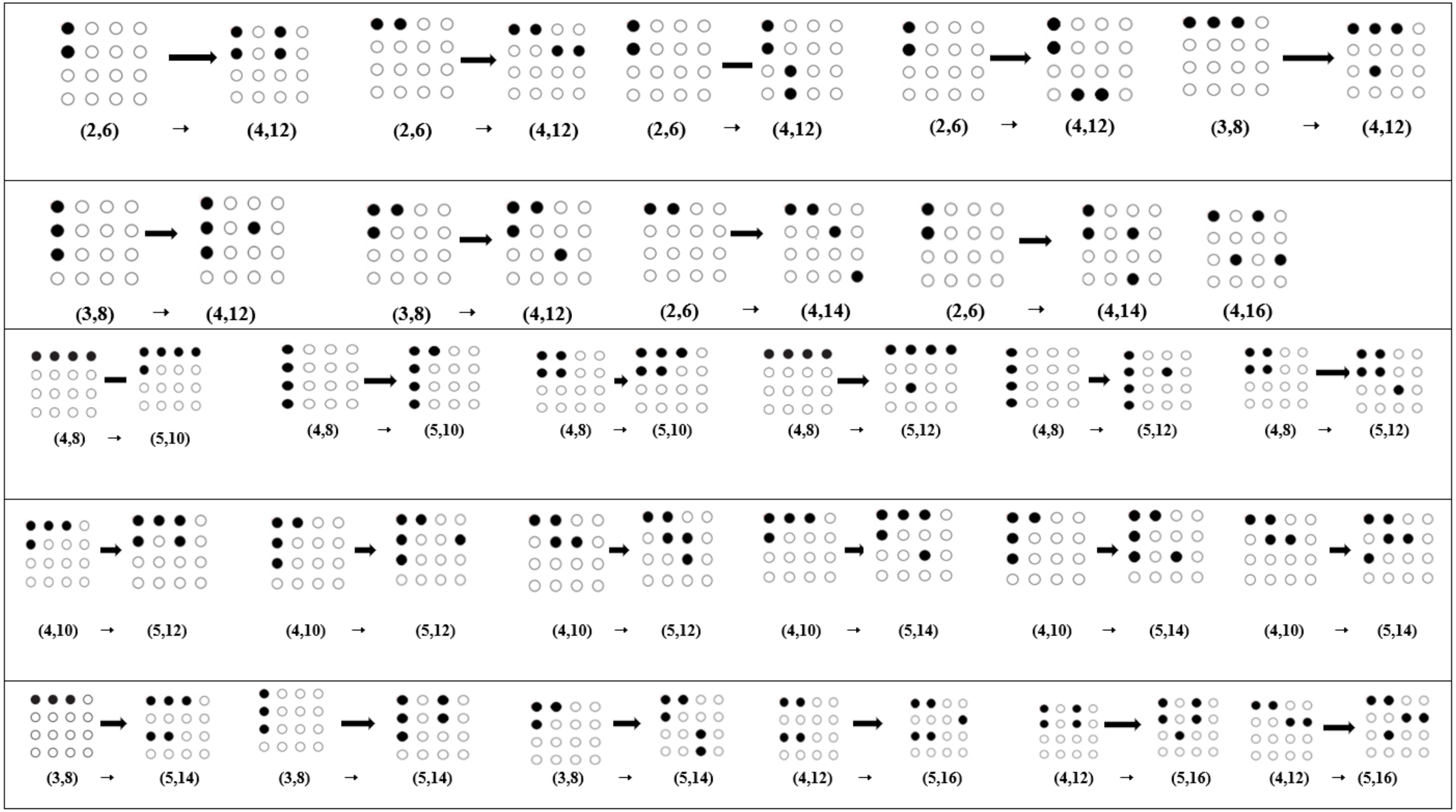

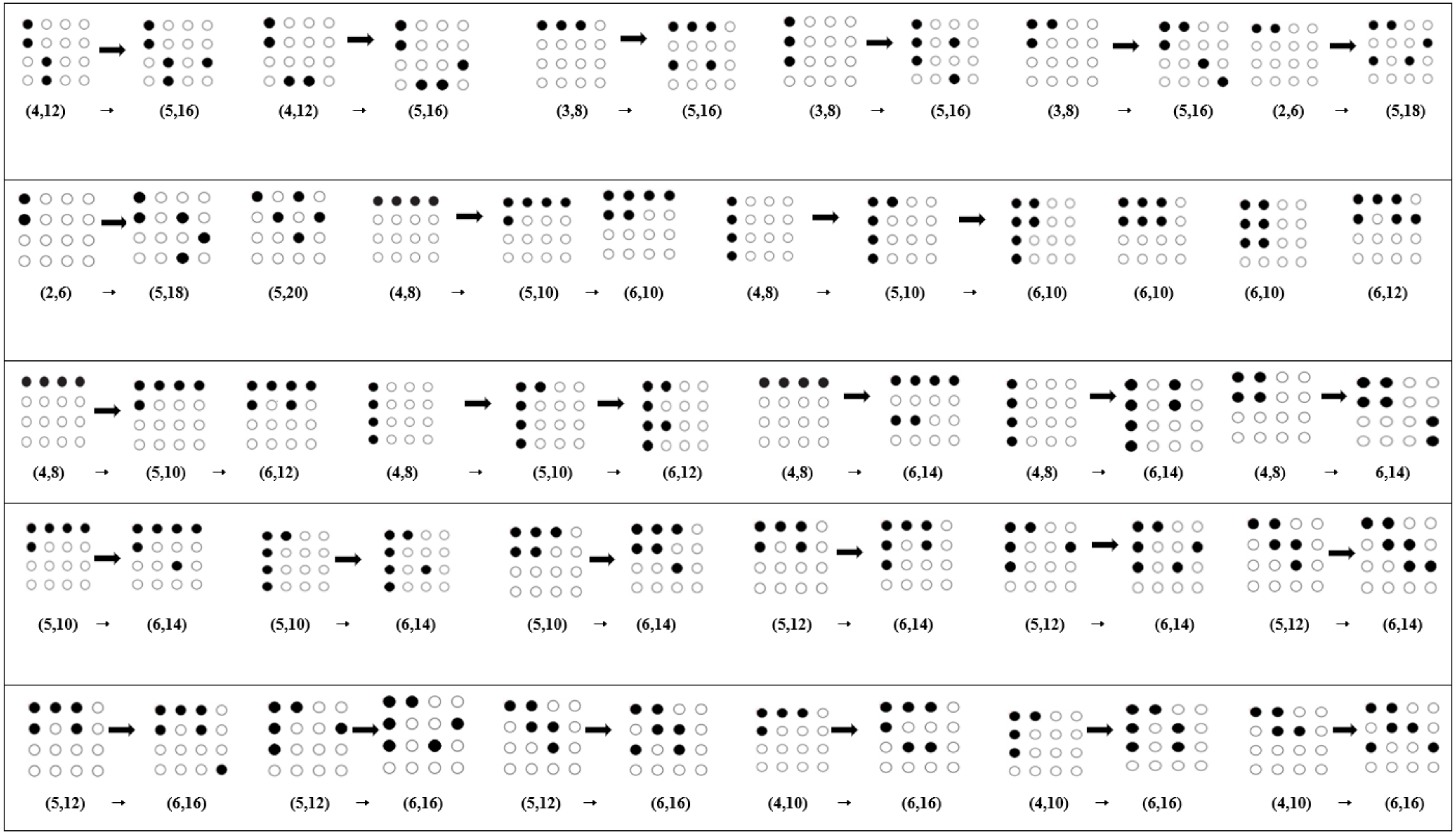

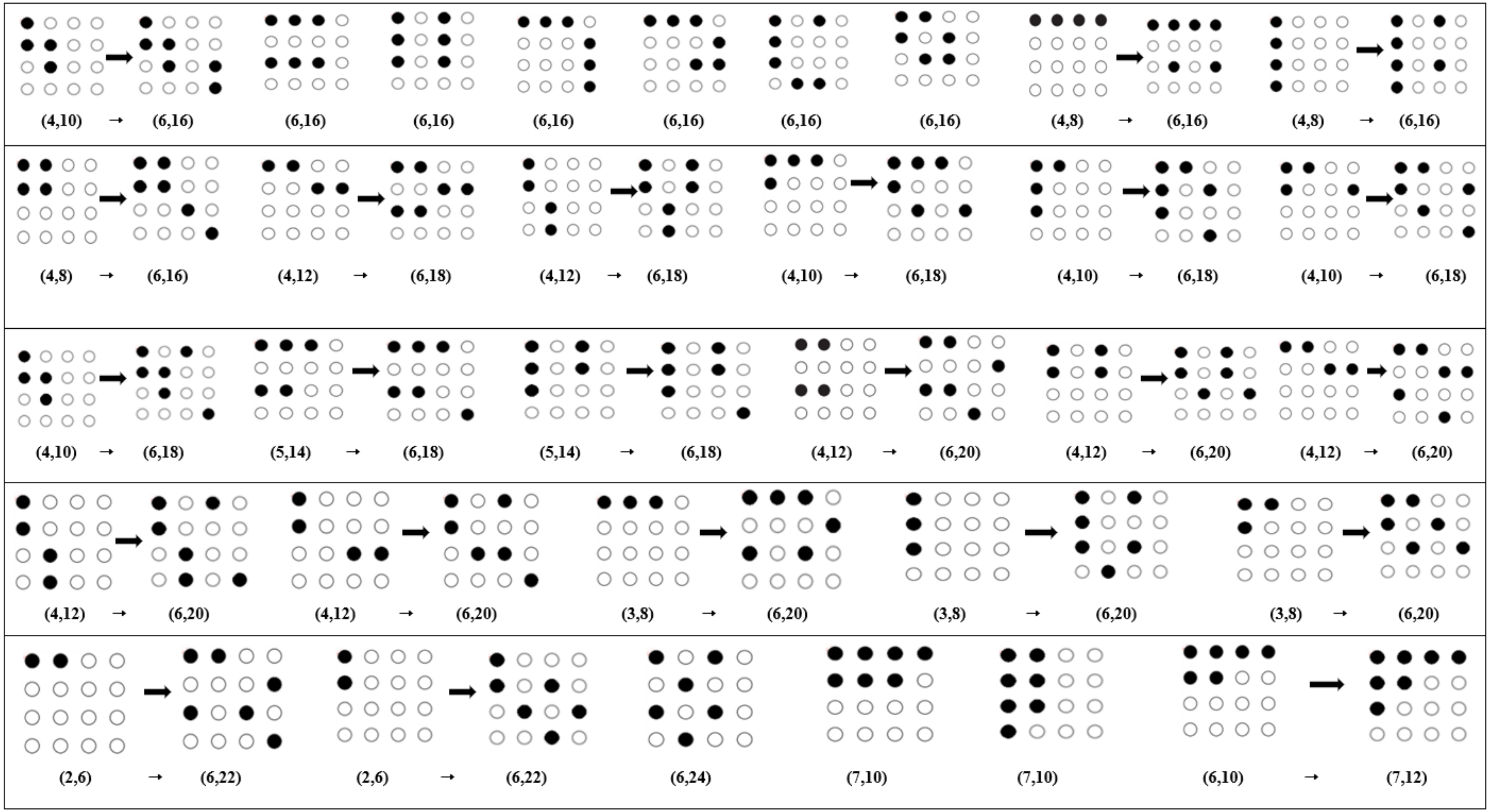

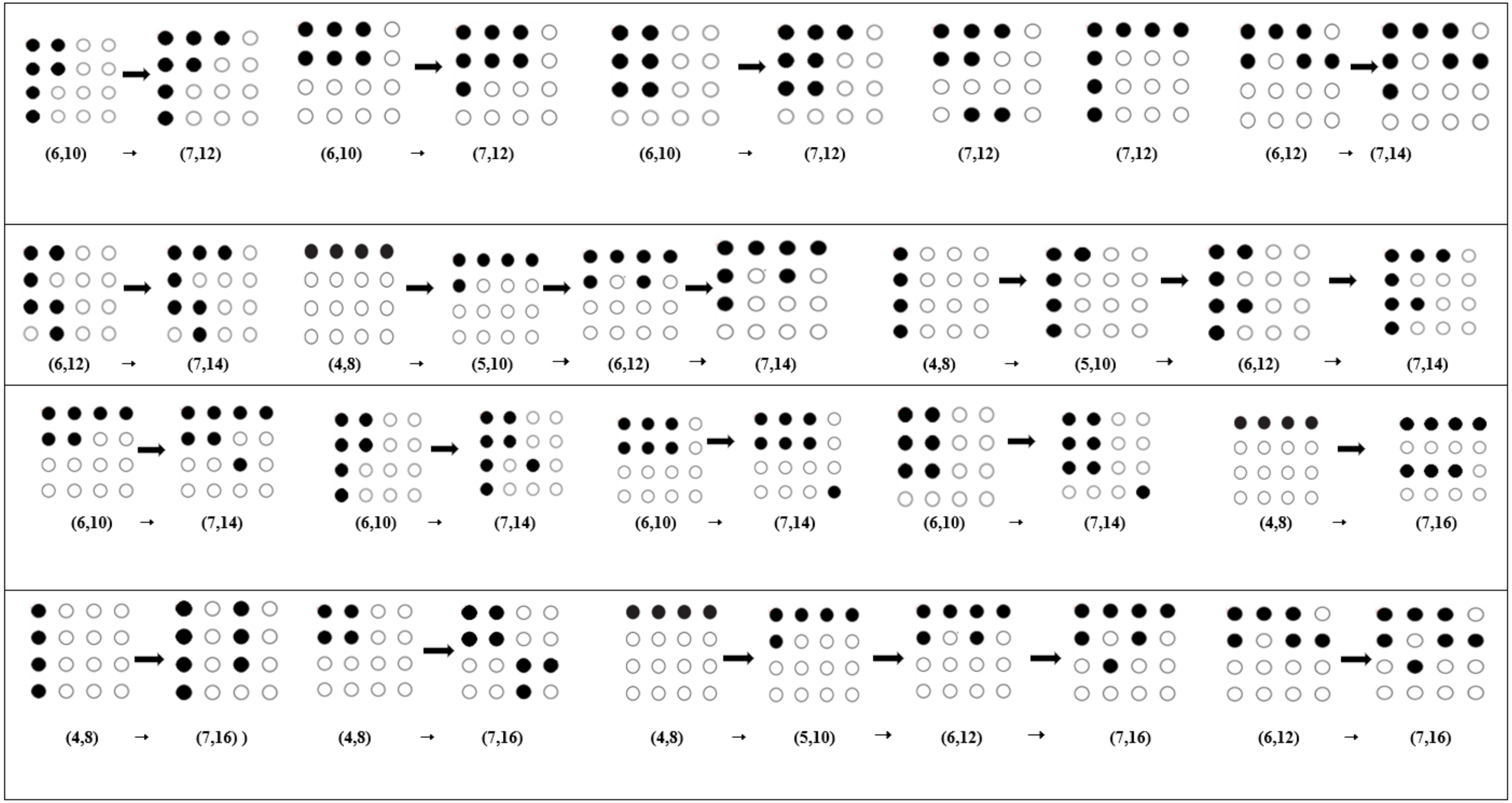
(6,10) → (7,12)
(6,10) → (7,12)
(6,10) → (7,12)
(7,12)
(7,12)
(6,12) → (7,14)
(6,12) → (7,14)
(4,8) → (5,10) → (6,12) → (7,14)
(4,8) → (5,10) → (6,12) → (7,14)
(6,10) → (7,14)
(6,10) → (7,14)
(6,10) → (7,14)
(6,10) → (7,14)
(4,8) → (7,16)
(4,8) → (7,16) )
(4,8) → (7,16)
(4,8) → (5,10) → (6,12) → (7,16)
(6,12) → (7,16)

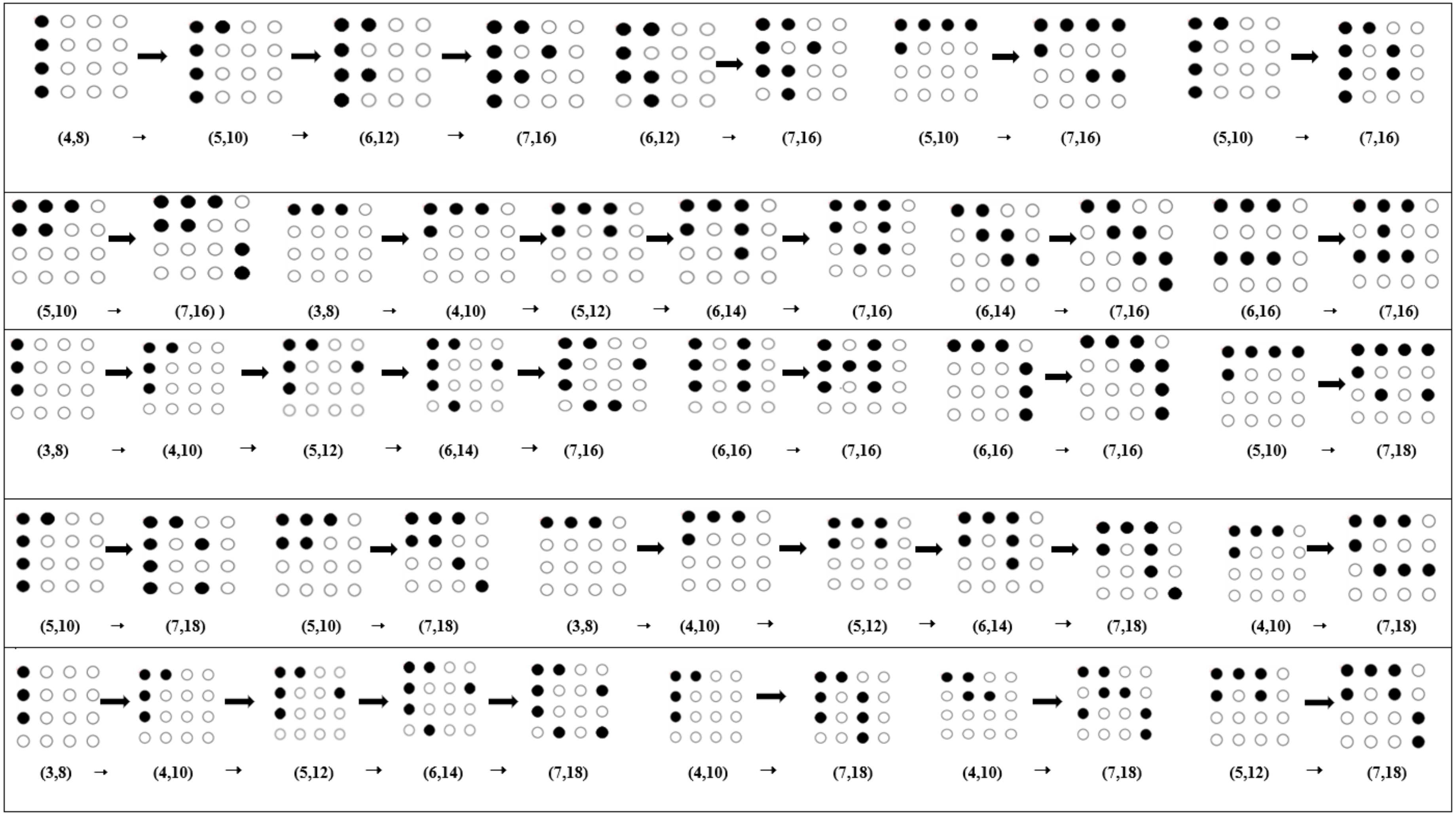
(4,8) → (5,10) → (6,12) → (7,16) (6,12) → (7,16) (5,10) → (7,16) (5,10) → (7,16)
(5,10) → (7,16) ) (3,8) → (4,10) → (5,12) → (6,14) → (7,16) (6,14) → (7,16) (6,16) → (7,16)
(3,8) → (4,10) → (5,12) → (6,14) → (7,16) (6,16) → (7,16) (6,16) → (7,16) (5,10) → (7,18)
(5,10) → (7,18) (5,10) → (7,18) (3,8) → (4,10) → (5,12) → (6,14) → (7,18) (4,10) → (7,18)
(3,8) → (4,10) → (5,12) → (6,14) → (7,18) (4,10) → (7,18) (4,10) → (7,18) (5,12) → (7,18)

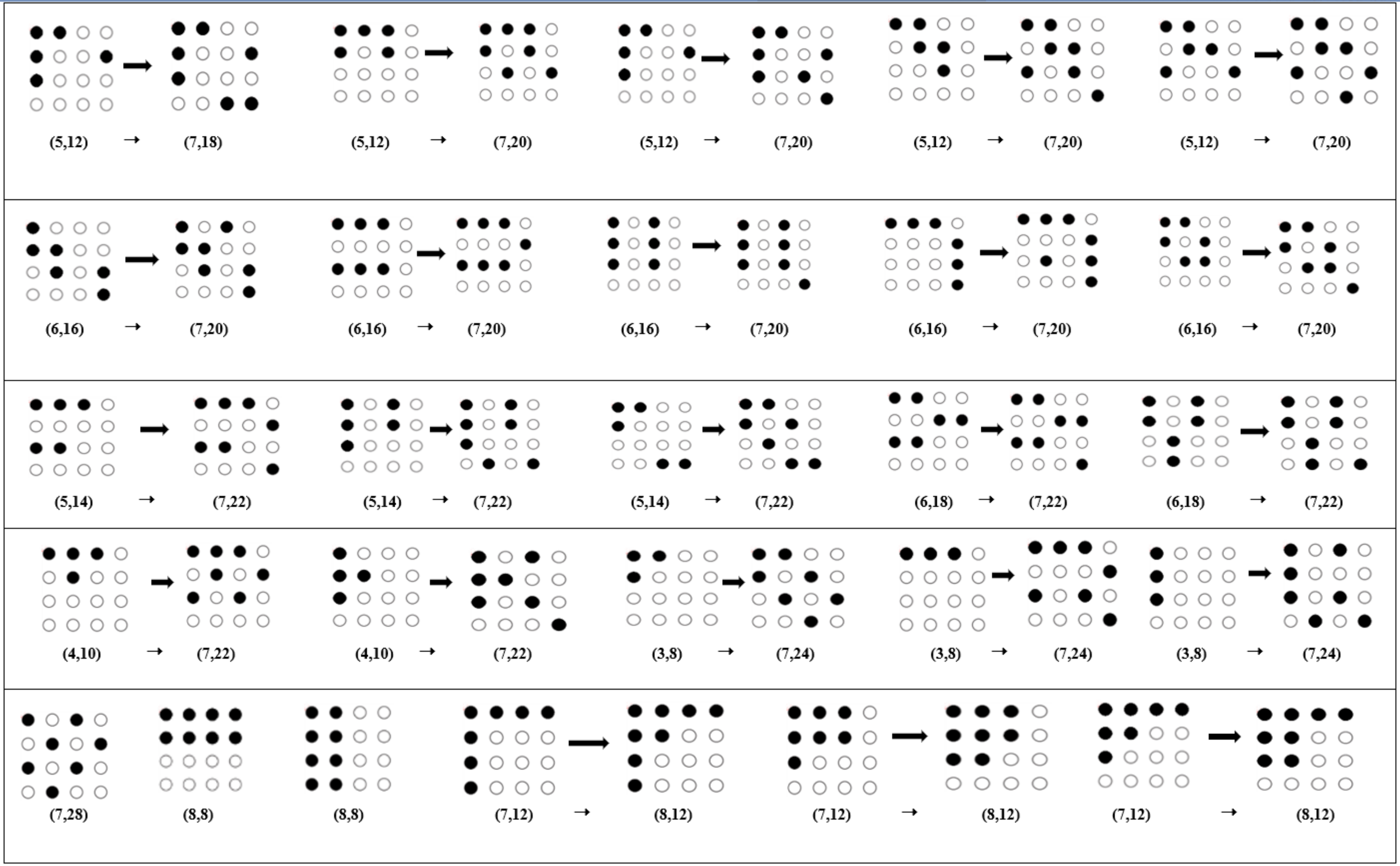

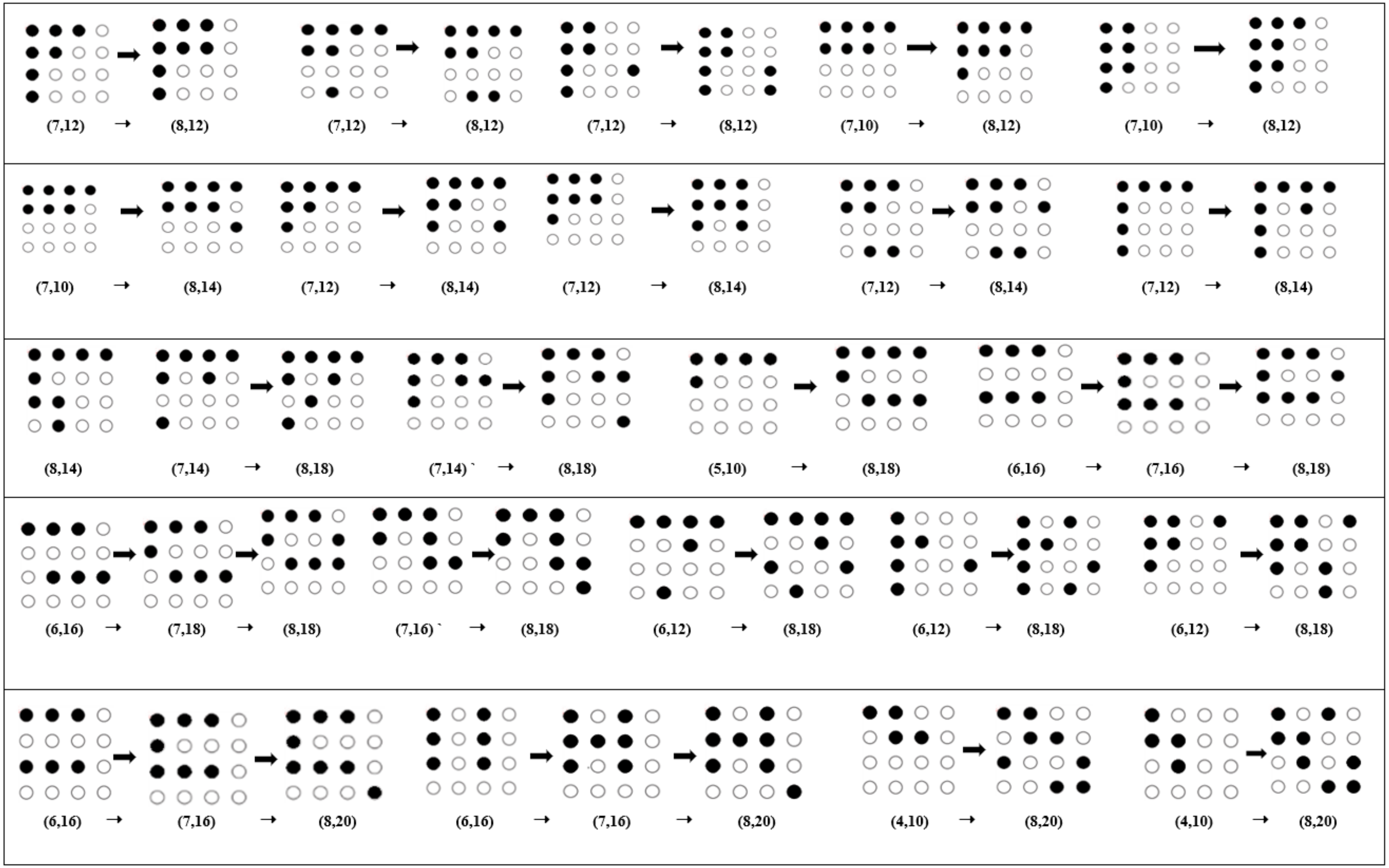

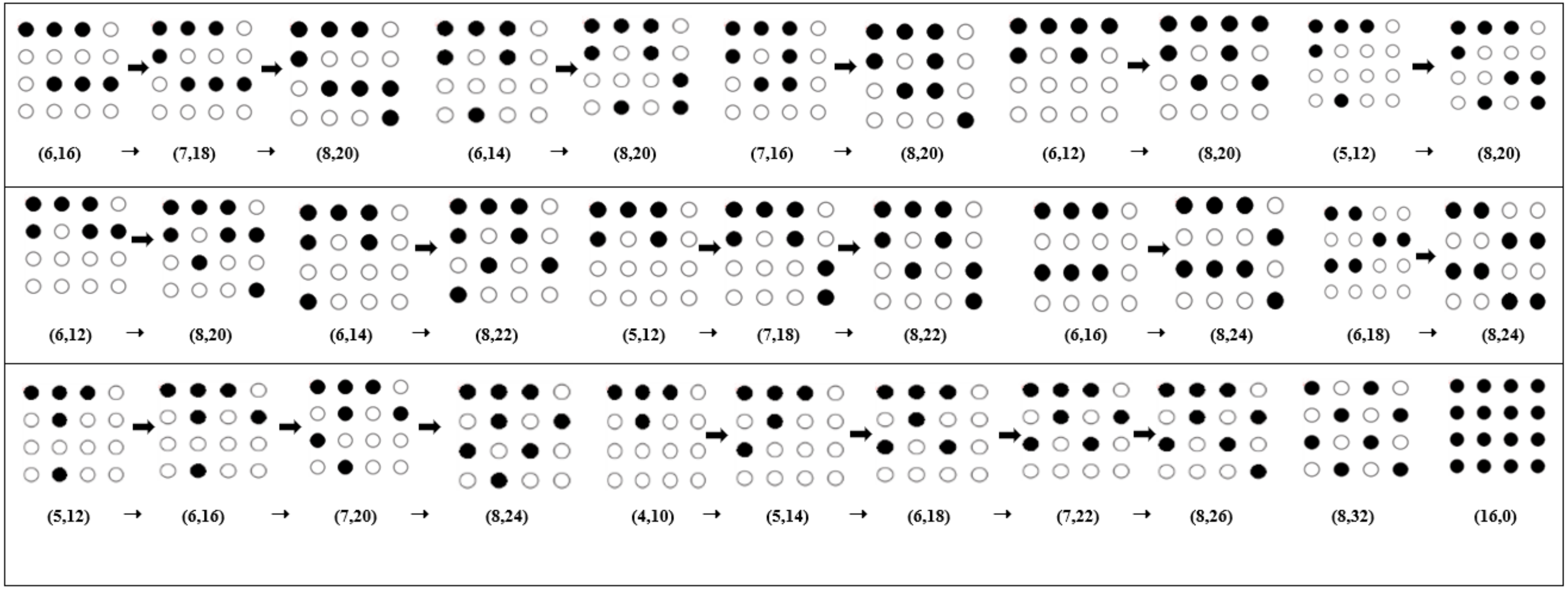

Table 1 provides a plausible representation of A(p,q) for different B-W edges ,arising from a chosen value of p. This analytical formula has been validated by considering each diagram with the generated B-W edges.

*Table 1*-Enumeration of the number of ways in which B-W edges (A(p,q) can arise for a square lattice of 16 sites. As a typical example, (6,10) →(7,12) illustrates the visualisation of an arrangement with '6' black sites and '10' black-white edges into another arrangement with '7' black sites and '12' black-white edges . Analogously for all other entries.

| A(p, q) | p | q | Number of ways of obtaining A(p,q) *vis a vis* the corresponding equation | Description of the arrangement |
|---|---|---|---|---|
| 1 | 0 | 0 | 1 | Only white sites |
| 16 | 1 | 4 | N | One non-adjacent B site |
| 32 | 2 | 6 | 2N | Pair of two adjacent B sites |
| 88 | 2 | 8 | $\left[\left\{2\sqrt{N}\left(\sqrt{N}-3\right)\left(1+\sqrt{N}\right)\right\}+\left\{\left(\sqrt{N}-1\right)N\right\}\right]$ | Two non- adjacent B sites |
| 96 | 3 | 8 | 6N | Pair of three adjacent B sites |
| 256 | 3 | 10 | $2N(N-8)$ | (2,6)→(3,10) |
| 208 | 3 | 12 | $\sqrt{N}\begin{bmatrix}\left\{2\left(\sqrt{N}-3\right)(N-8)\right\}+\left\{\left(\sqrt{N}-1\right)(N-10)\right\}\\+\left\{\left(4\sqrt{N}-7\right)+(\sqrt{N}-1)^{2}\right\}\end{bmatrix}$ | Three non-adjacent B sites |
| 24 | 4 | 8 | $\sqrt{N}(\sqrt{N}+2)$ | Pair of four adjacent B sites |
| 256 | 4 | 10 | 16N | (3,8)→(4,10) |
| 736 | 4 | 12 | $N\left[\left\{4\left(\sqrt{N}-3\right)\right\}+6+\{(6)(N-10)\}\right]$ | 1.(2,6)→(4,12) 2. (3,8)→(4,12) |
| 576 | 4 | 14 | $4N\sum_{x=0}^{2}[(N-11)-(xd)]$ | (2,6)→(4,14) |
| 228 | 4 | 16 | $\left[\begin{array}{c}\sqrt{N}\{4(\sqrt{N}-3)(3N-40)+2+(3\sqrt{N}-5)(\sqrt{N}-1)\} \\ +8(\sqrt{N}-3)\end{array}\right]$ | Four non-adjacent B sites |
| 192 | 5 | 10 | $(8)\left(\sqrt{N}\right)\left(2+\sqrt{N}\right)$ | (4,8)→(5,10) |
| 688 | 5 | 12 | $\left[\begin{array}{c}\left\{\left(N+2\sqrt{N}\right)(N-12)\right\}+N\sum_{i=1}^{2(p-2)} i \\ +\{N(12)\}+4N\end{array}\right]$ | 1. (4,8)→(5,12)<br>2. (4,10)→(5,12) |
| 1664 | 5 | 14 | $2N\left(8N+\sqrt{N}-80\right)$ | 1. (4,10)→(5,14) 2. (3,8)→(5,14)<br>s |
| 1248 | 5 | 16 | $\left[\{N(2N+38)\}+\left\{32\sqrt{N}\left(\sqrt{N}-3\right)\right\}\right]$ | 1.(4,12)→(5,16) 2. (3,8)→(5,16) |
| 448 | 5 | 18 | 4N (N-9) | (2,6)→(5,18) |
| 128 | 5 | 20 | $\left(\sqrt{N}-3\right)\left[\left\{\left(6\sqrt{N}-12\right)(N-12)\right\}+20\sqrt{N}\right]$ | Five non-adjacent B sites |
| 96 | 6 | 10 | 6N | 1.(4,8)→(5,10)→(6,10)<br>2. Two superimposing pairs of three adjacent B sites |
| 704 | 6 | 12 | $[\{4(2N-25)\left(2\sqrt{N}+N\right)\}+2N]$ | 1.Two pairs of three adjacent B sites such that no edges with four adjacent B sites is formed.<br>2.(4,8)→(5,10)→(6,12)<br>s |
| 1824 | 6 | 14 | $\left[78N+\left\{24\left(2\sqrt{N}+N\right)\right\}\right]$ | 1.(4,8)→(6,14) 2. (5,10)→(6,14)<br>3.(5,12)→(6,14) |
| 2928 | 6 | 16 | $\left[\left\{164+4\left(\sqrt{N}-3\right)\right\}N+\left\{44+4\sqrt{N}\right\}\sqrt{N}\right]$ | 1. (5,12)→(6,16) 2. (4,10)→(6,16)<br>3. Two non-adjacent pairs of three contiguous B sites<br>4. (4,8)→(6,16) |
| 1568 | 6 | 18 | 98N | 1. (4,12)→(6,18) 2. (4,10)→(6,18) 3. (5,14)→(6,18) |
| 768 | 6 | 20 | $\left[\left\{24\sqrt{N}\left(\sqrt{N}-3\right)\right\}+\{(42)N\}\right]$ | 1.(4,12)→(6,20) 2. (3,8)→(6,20) |
| 64 | 6 | 22 | 4N | (2,6)→(6,22) |
| 56 | 6 | 24 | $2\left(13\sqrt{N}-24\right)$ | Six non-adjacent B sites. |
| 64 | 7 | 10 | 4N | Adjacent pair of four adjacent B sites (4,8) and pair of three adjacent B sites (3,8) |
| 624 | 7 | 12 | $2\sqrt{N}\left(19\sqrt{N}+2\right)$ | 1.(6,10)→(7,12)<br>2.Pair of three adjacent B sites in rows and two pairs of two adjacent B sites<br>3.Two pairs of four adjacent B sites interpenetrating each other |
| 1920 | 7 | 14 | $8\sqrt{N}\left(16+11\sqrt{N}\right)$ | 1. (6,12)→(7,14)<br>2. (6,10)→(7,14) |
| 3680 | 7 | 16 | $2\sqrt{N}\left(125\sqrt{N}-40\right)$ | 1. (4,8)→(7,16) 2. (6,12)→(7,16)<br>3. (5,10)→(7,16) 4. (6,14)→(7,16)<br>5. (6,16)→(7,16) |
| 3136 | 7 | 18 | $16\sqrt{N}\left(12\sqrt{N}+1\right)$ | 1.(5,10)→(7,18) 2.(6,14)→(7,18)<br>3. (4,10)→(7,18) 4. (5,12)→(7,18) |
| 1392 | 7 | 20 | 87N | 1. (5,12)→(7,20) 2. (4,10)→(6,16)→(7,20)<br>3. (4,10)→(6,16)→(7,20)<br>4. (6,16)→(7,20) |
| 512 | 7 | 22 | 32N | 1. (5,14)→(7,22) 2.(6,18)→(7,22)<br>3. (4,10)→(7,22) |
| 96 | 7 | 24 | 6N | Three adjacent B sites (3,8) and four non-adjacent B sites (4,16) |
| 0 | 7 | 26 | 0 | It is not possible to have 7 B sites yielding 26 edges in a square lattice of 16 vertices. |
| 16 | 7 | 28 | $4\sqrt{N}\left(\sqrt{N}-3\right)$ | Seven non-adjacent B sites |
| 8 | 8 | 8 | $2\sqrt{N}$ | Two superimposing pairs of four adjacent B sites along adjacent rows and columns |
| 0 | 8 | 10 | 0 | This configuration is not possible. |
| 768 | 8 | 12 | $4\sqrt{N}\left(4+11\sqrt{N}\right)$ | 1. (7,12)→ (8,12) 2.(7,10)→(8,12) |
| 1600 | 8 | 14 | $16\sqrt{N}\left(1+6\sqrt{N}\right)$ | 1.(7,10)→(8,14) 2. (7,12)→(8,14)<br>3.Two interpenetrating pairs of four adjacent B sites (along rows or columns) and pair of three adjacent B sites |
| 4356 | 8 | 16 | $\left[2\sqrt{N}(98 + 112\sqrt{N}) - 12\right]$ | 1. (4,8)→ (5,10) → (6,12) → (7,14) → (8,16)<br>2.(4,8)→(8,16)<br>3.A(6,10) → (8,16)<br>4. A(7,12)→ (8,16)<br>5.A(6,12) → (7,14) → (8,16)<br>6. A(6,10) → (7,12) → (8,16)<br>7. A(6,16) → (7,16) → (8,16)<br>8. A(6,14) → (7,16) → (8,16) |
| 3264 | 8 | 18 | $\left[132N+\left\{8\sqrt{N}\left(16+20(\sqrt{N}-3\right)\right\}\right]$ | 1(7,14)→(8,18)<br>2.(5,10)→(8,18) 3.(6,12)→(8,18)<br>4.(6,16)→(7,16)→(8,18)<br>5.(6,16)→(7,18)→(8,18 ) 6.(7,16)→(8,18) |
| 2112 | 8 | 20 | $\left[102N+\left\{8\sqrt{N}\left(4+11(\sqrt{N}-3\right)\right\}\right]$ | 1.(6,16)→(7,16)→(8,20) 2.(4,10)→(8,20)<br>3.(6,16)→(7,18)→(8,20) 4. (6,14)→(8,20)<br>5.(7,16)→(8,20) 6.(6,12)→(8,20) |
| | | | | 7.(5,12)→(8,20) |
| 576 | 8 | 22 | 36N | 1.(6,14)→(8,22)<br>2. (5,12)→(7,18)→(8,22) |
| 120 | 8 | 24 | $6\sqrt{N}\left(2\sqrt{N}\text{-}3\right)$ | 1. (6,16)→(8,24) 2. (6,18)→(8,24)<br>3.(5,12)→(6,16)→(7,20)→(8,24) |
| 64 | 8 | 26 | 4N | (4,10)→(5,14)→(6,18)→(7,22)→(8,26) |
| 0 | 8 | 28 | 0 | This configuration- not possible |
| 0 | 8 | 30 | 0 | This configuration- not possible |
| 2 | 8 | 32 | $2\left(\sqrt{N}-3\right)$ | Eight non- adjacent black sites |
| 1 | 16 | 0 | 1 | Only black sites and hence no B-W edges |

**2.2 Diagrammatic Analysis of A(8,16)**

While the calculation of a few entries of Table 1 is straight-forward, several others are indeed tedious for deducing the analytical closed-form expression. One such example is provided by the enumeration of the number of ways in which eight black sites and sixteen black-white edges can arise. By a systematic methodology, this number is deduced as 4356. The methodology is provided in Fig 2.Such computations have not hitherto-been obtained using analytical reasoning.

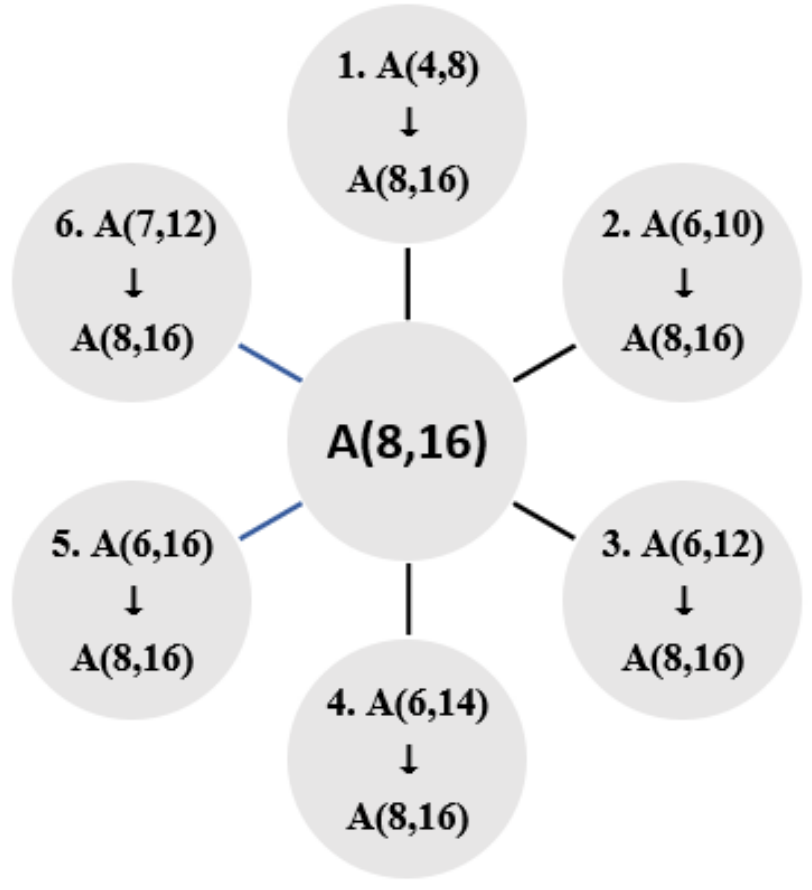

**Figure 2:** A hierarchical methodology of deducing A(8,16) from earlier configurations

The various transformations depicted in Fig 2 are briefly described below:

1. A(4,8)→A(8,16) indicates transformation of a configuration having four B sites and eight corresponding B-W edges into a configuration consisting of eight B sites and sixteen corresponding B-W edges. This conversion can occur in two ways.

(a) Successive substitution of four B sites , ensuring that no repetition occurs and the resulting expression is $116\sqrt{N}(2+\sqrt{N})$.

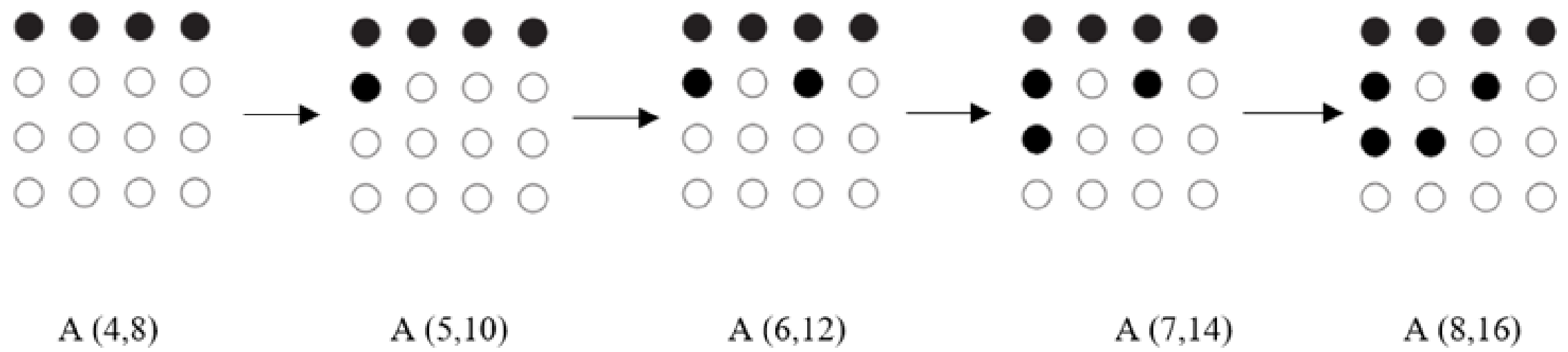

**Figure 2a:** Transformation of A(4,8) to A(8,16)

(b)This method involves substitution of two non-adjacent pairs of four mutually adjacent B sites as shown below: The total number of ways by which this arrangement can arise is $6\sqrt{N} - 12$.

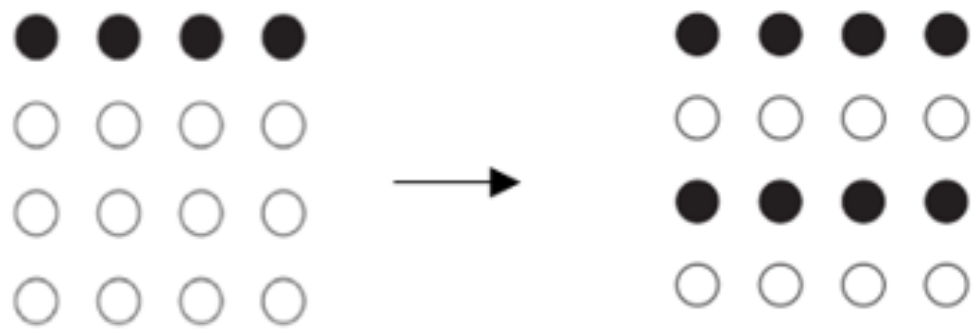

**Figure 2b:** Transformation of A(4,8) to A(8,16)

2. A(6,10)→A(8,16) indicates hierarchical transformation of a configuration having 6 B sites and 10 corresponding B-W edges into a configuration consisting of 8 B sites and 16 B-W edges. This conversion also can occur in two ways.

(a) By successive addition of one adjacent and one non-adjacent B site in the diagram pertaining to A(6,10) for obtaining the entries of A(8,16) viz $48N$ .

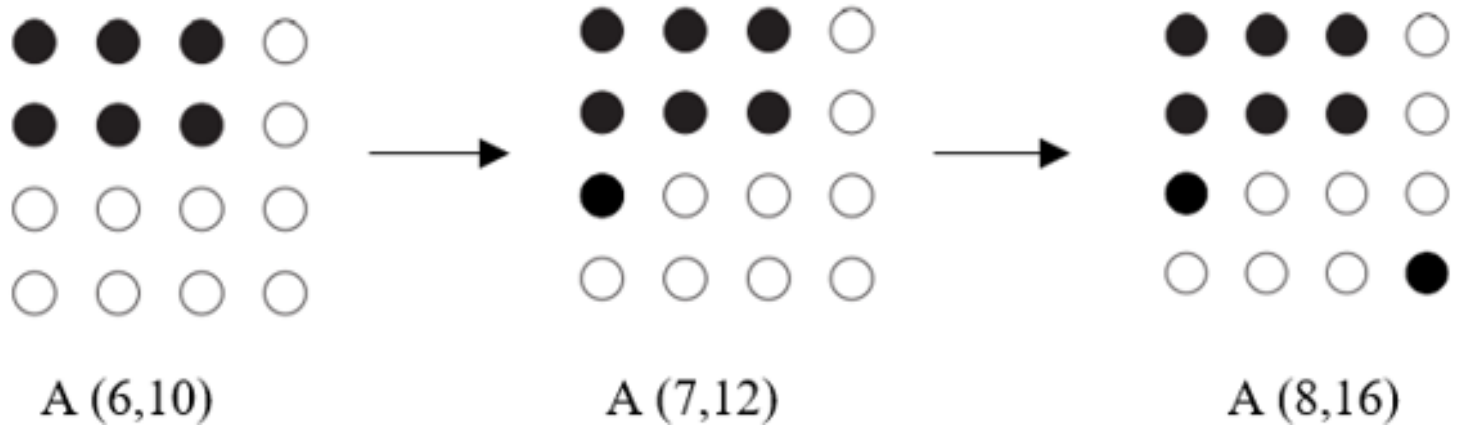

**Figure 2c:** Transformation of A(6,10) to A(8,16)

(b) The substitution of two non-adjacent pairs of two adjacent B sites in A(6,10) yields $6N$ desired configurations.

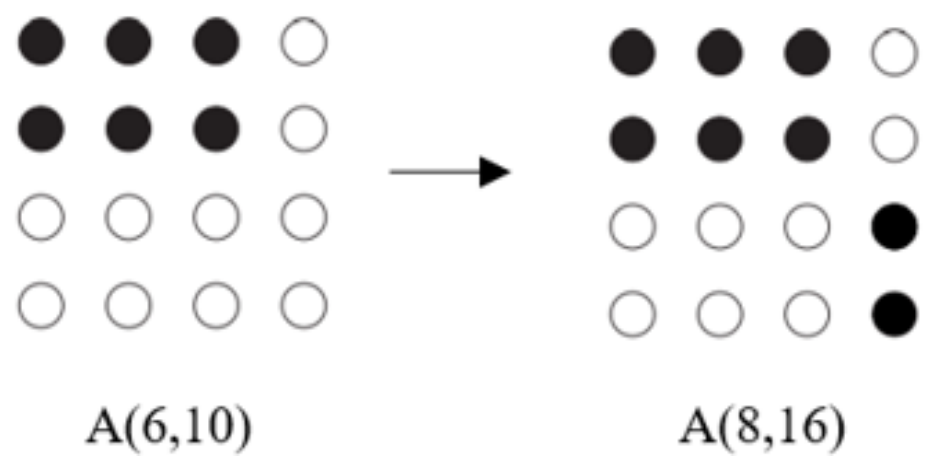

**Figure 2d:** Transformation of A(6,10) to A(8,16)

3. A(6,12)→A(8,16) transformation occurs through successive addition of two B sites without any repetitions and this leads to $30N$ configurations with N =16.

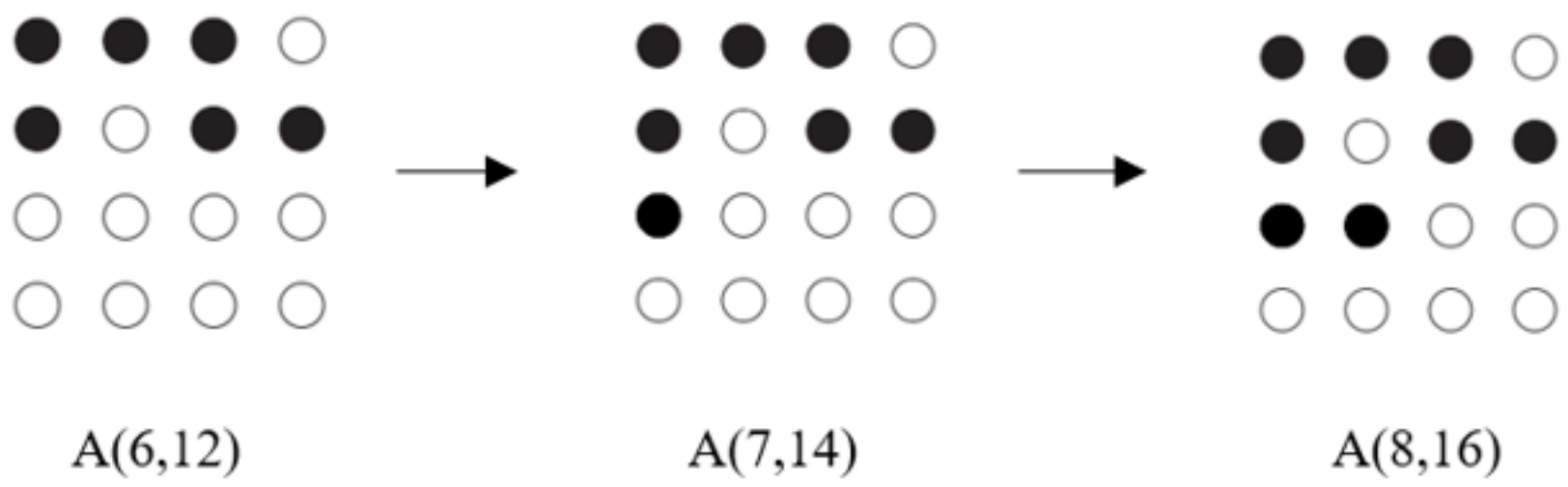

**Figure 2e:** Transformation of A(6,12) to A(8,16)

4. A(6,14)→A(8,16) involves successive substitution of two B sites , without repetition and yields $2\sqrt{N}$ arrangements

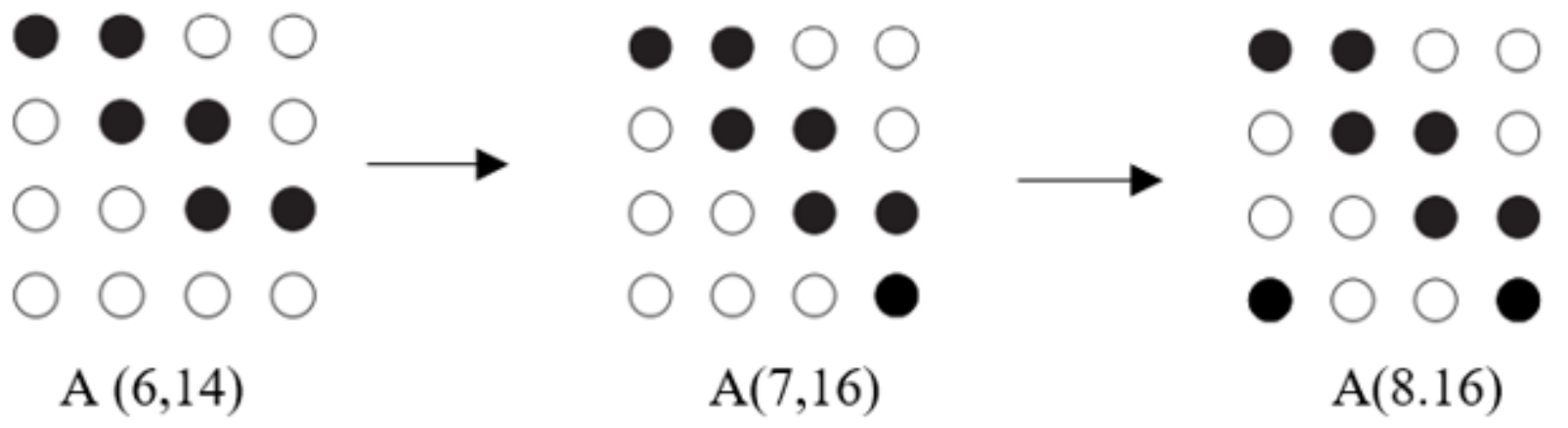

**Figure 2f:** Transformation of A(6,14) to A(8,16)

5. A(6,16)→A(8,16) involves successive substitution of two B sites as in the above case which leads to $2\sqrt{N}\left(11\sqrt{N}-24\right)$ configurations.

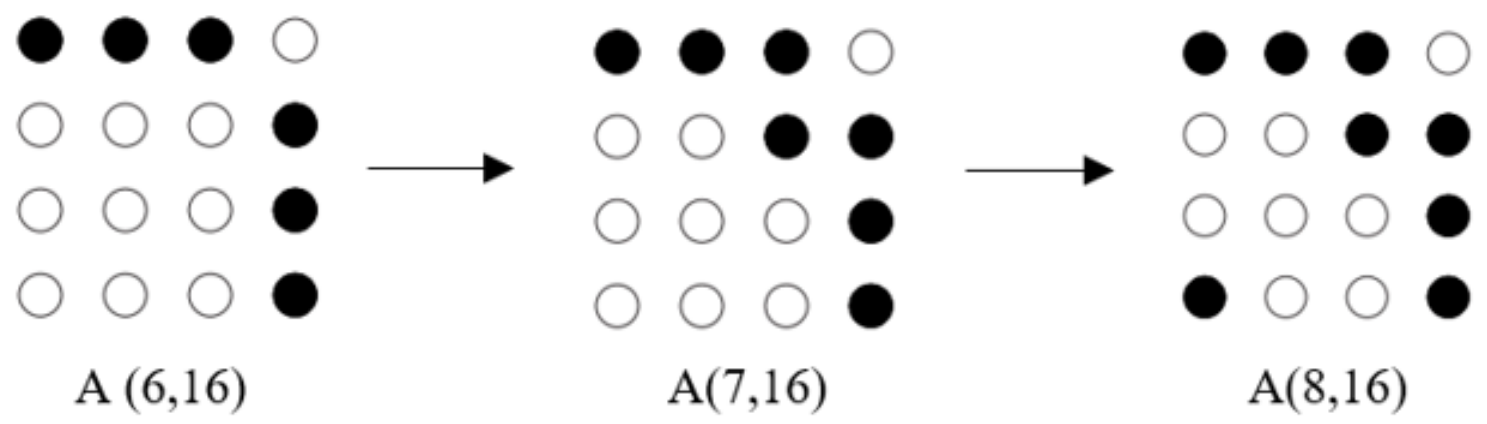

**Figure 2g:** Transformation of A(6,16) to A(8,16)

**6.** A(7,12)→A(8,16) involves substitution of one non-adjacent B site in the lower configuration which leads to $2\sqrt{N}\left(2+\sqrt{N}\right)$ configurations.

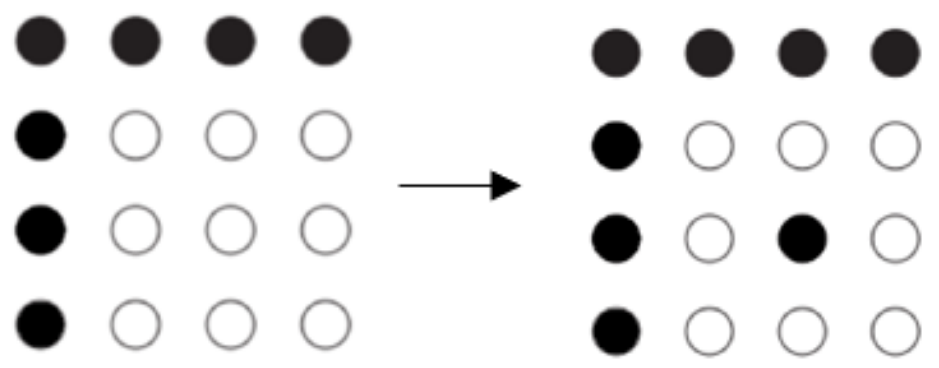

**Figure 2h:** Transformation of A(7,12) to A(8,16)

## 2.4 Expression for the canonical partition function

It is well known [24] that the canonical partition function can be written in the parlance of combinatorics as

$$Q(x,y) = \sum_{p,q} A(p,q) x^{2(N-q)} y^{N-2p} \tag{3}$$

where $x = e^{J/kT}$ and $y = e^{H/kT}$. Upon substituting all the entries of A(p,q) of Table 1, we obtain the partition function for a square lattice of 16 sites as

$$
\begin{aligned}
lnQ(H,J) = ln\Big[&2e^{\frac{32J}{kT}}cosh\left(\frac{16H}{kT}\right) + 32e^{\frac{24J}{kT}}cosh\left(\frac{14H}{kT}\right) + 64e^{\frac{20J}{kT}}cosh\left(\frac{12H}{kT}\right) \\
&+ 8e^{\frac{16J}{kT}}\left[22cosh\left(\frac{12H}{kT}\right) + 24cosh\left(\frac{10H}{kT}\right) + 6cosh\left(\frac{8H}{kT}\right) + 1\right] \\
&+ 64e^{\frac{12J}{kT}}\left[8cosh\left(\frac{10H}{kT}\right) + 8cosh\left(\frac{8H}{kT}\right) + 6cosh\left(\frac{6H}{kT}\right) + 3cosh\left(\frac{4H}{kT}\right) + 2cosh\left(\frac{2H}{kT}\right)\right] \\
&+ 32e^{\frac{8J}{kT}}\left[13cosh\left(\frac{10H}{kT}\right) + 46cosh\left(\frac{8H}{kT}\right) + 43cosh\left(\frac{6H}{kT}\right) + 44cosh\left(\frac{4H}{kT}\right) + 39cosh\left(\frac{2H}{kT}\right) + 24\right] \\
&+ 64e^{\frac{4J}{kT}}\left[18cosh\left(\frac{8H}{kT}\right) + 52cosh\left(\frac{6H}{kT}\right) + 57cosh\left(\frac{4H}{kT}\right) + 60cosh\left(\frac{2H}{kT}\right) + 25\right] \\
&+ 4\left[114cosh\left(\frac{8H}{kT}\right) + 624cosh\left(\frac{6H}{kT}\right) + 1{,}464cosh\left(\frac{4H}{kT}\right) + 1840cosh\left(\frac{2H}{kT}\right) + 1{,}089\right] \\
&+ 64e^{\frac{-4J}{kT}}\left[14cosh\left(\frac{6H}{kT}\right) + 49cosh\left(\frac{4H}{kT}\right) + 98cosh\left(\frac{2H}{kT}\right) + 51\right] \\
&+ 32e^{\frac{-8J}{kT}}\left[8cosh\left(\frac{6H}{kT}\right) + 48cosh\left(\frac{4H}{kT}\right) + 87cosh\left(\frac{2H}{kT}\right) + 66\right] \\
&+ 64e^{\frac{-12J}{kT}}\left[2cosh\left(\frac{4H}{kT}\right) + 16cosh\left(\frac{2H}{kT}\right) + 9\right] + 8e^{\frac{-16J}{kT}}\left[14cosh\left(\frac{4H}{kT}\right) + 24cosh\left(\frac{2H}{kT}\right) + 15\right] + 64e^{\frac{-20J}{kT}} \\
&+ 32e^{\frac{-24J}{kT}}cosh\left(\frac{2H}{kT}\right) + 2e^{\frac{-32J}{kT}}\Big] \qquad (4)
\end{aligned}
$$

Although the above eqn appears algebraically complex, simplifications are indeed possible, after ascertaining the importance of each Boltzmann factor. By numerical manipulation of the above equation for the partition function, the internal energy, specific heat, magnetization and susceptibility can also be derived, albeit for N =16. An alternate form of eqn (4) has been provided earlier[25].

## 3. Magnetization for finite sites in non-zero magnetic field

It is well known[21] that magnetization can be defined in terms of the Helmholtz free energy ( A) as

$$A = -kT\, logQ \quad (5)$$

$$M_T = \left(\frac{\partial}{\partial H} logQ\right)_T \quad (6)$$

Although the earlier canonical partition function for 16 sites is exact, its utility is limited in so far as it cannot predict the onset of critical phenomena , in view of the analysis being restricted to finite sites. An obvious method of obviating this limitation consists in extending this to various square lattices of larger sizes and extrapolating the result to infinite sites( 1/N tending to zero). This approach while yielding numerical values cannot provide an analytical expression . Consequently, we postulate a semi-empirical eqn for magnetization whose validity is shown using the numerical computation of magnetization at different magnetic fields.

### 3.2 Magnetization for infinite sites in a non-zero magnetic field

It has been well-recognised that a rigorous exact formulation of the magnetization in the presence of a finite magnetic field is indeed a formidable task [1].However, there are several methods by which quantitatively exact expressions can be deduced viz(i) extrapolation to infinite sites using Pade' approximation [12] (ii) Enumeration strategies for infinite sites using the entries of Table 1 and (iii) finite size scaling analysis [26].

We have employed a semi-empirical approach to obtain a closed form expression for the partition function and will be discussed in a subsequent communication. Here, we provide the result pertaining to the field-dependent magnetization as obtained from the partition function *viz:*

$$M = \left[(M_{Ons})^{15} + sinh^4\left(\frac{2J}{kT}\right) tanh\left(\frac{H}{kT}\right) + \left(1 - \left\{sinh^4\left(\frac{2J}{kT}\right)\right\}\right) tanh^{15}\left(\frac{H}{kT}\right)\right]^{\frac{1}{15}} \qquad (7)$$

where $M_{Ons}$ denotes the equation for the spontaneous magnetization as deduced by Onsager [5]

$$M_{ons} = \left[1 - \left\{ sinh^{-4}\left(\frac{2J}{kT}\right)\right\}\right]^{\frac{1}{8}} \qquad (8)$$

Furthermore, when the nearest neighbour interaction energy ($J$) is zero, eqn (7) leads to

$$M = tanh(H/kT) \qquad (9)$$

as anticipated. For the non-zero magnetic field, the critical exponents are $\delta$ *and* $\gamma$, the former being associated with the magnetic field dependence of the magnetization at the critical temperature and the latter with the zero-field susceptibility. At the critical temperature, it follows from eqn (7) that M ~ $H^{1/15}$ for small values of H.

Fig 3 depicts the variation of M with H/kT for T/Tc >1 and T/Tc <1.Fig 4 depicts the dependence of M with dimensionless magnetic field for various J/kT values**.**

The field-dependent susceptibility is given by

$$\chi = \left(\frac{\partial M}{\partial H}\right)_T \sim \frac{1}{15kTM^{14}} \qquad (10)$$

At H =0, the susceptibility eqn follows as

$$\left(\frac{\partial M}{\partial H}\right)_{H=0} \sim \frac{1}{kT\left(1-\frac{T}{T_C}\right)^{\frac{7}{4}}} \tag{11}$$

The quantitative validity for eqn (7) may be inferred from the following : (i) when H=0, eqn (7) reduces to the well-known eqn for the spontaneous magnetization and thereby the correct exponent β for the temperature dependence;(ii) consistency with the exponent γ (7/4) for the zero-field susceptibility from eqn (11) and (iii) reproduction of the exponent δ for the magnetization at the critical temperature $T_c$ from eqn (7) The critical exponents $\alpha$ and $\beta$ pertaining to the specific heat at zero magnetic field will also follow from the Onsager's eqn for the spontaneous magnetization.

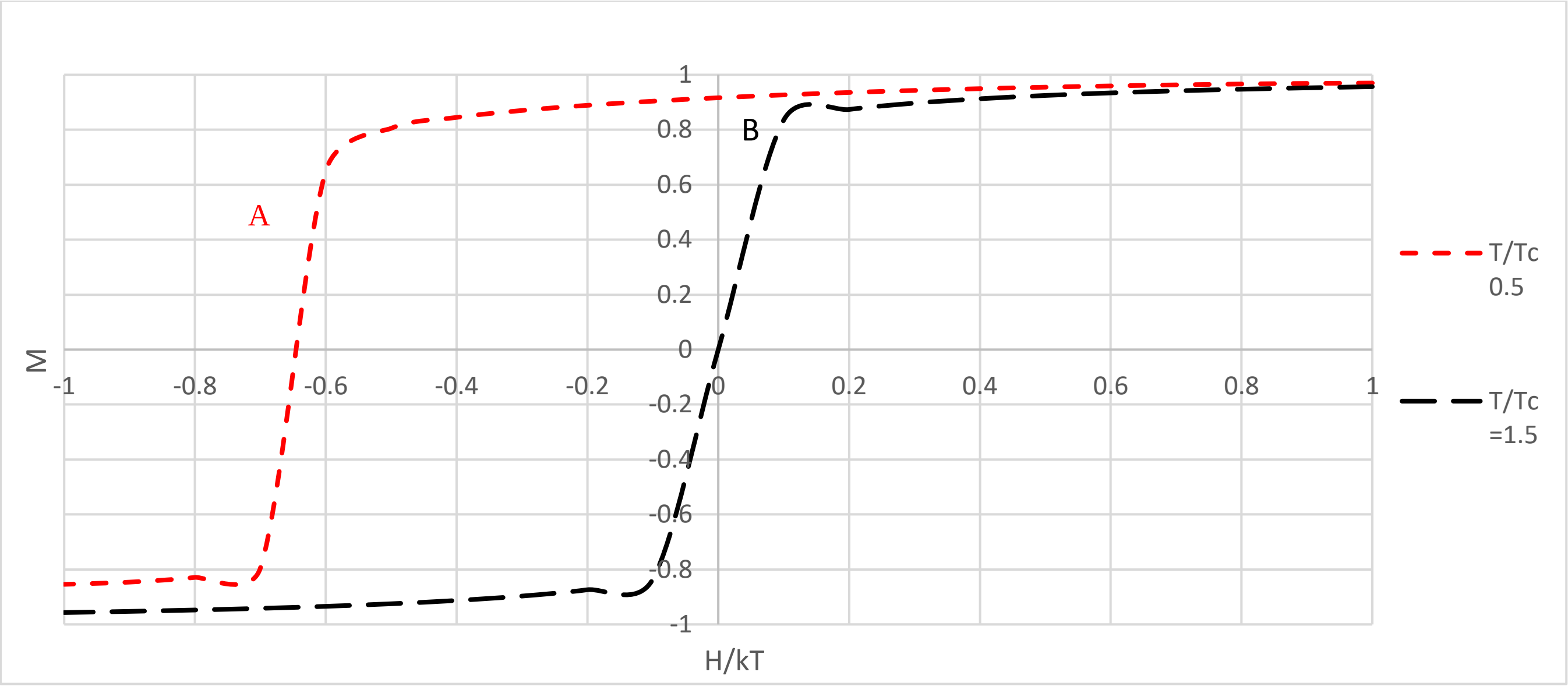

**Figure 3:** Dependence of the magnetization on dimensionless magnetic field for two temperatures. (A) T/Tc =0.5 and (B) T/Tc = 1.5 using eqn (7)

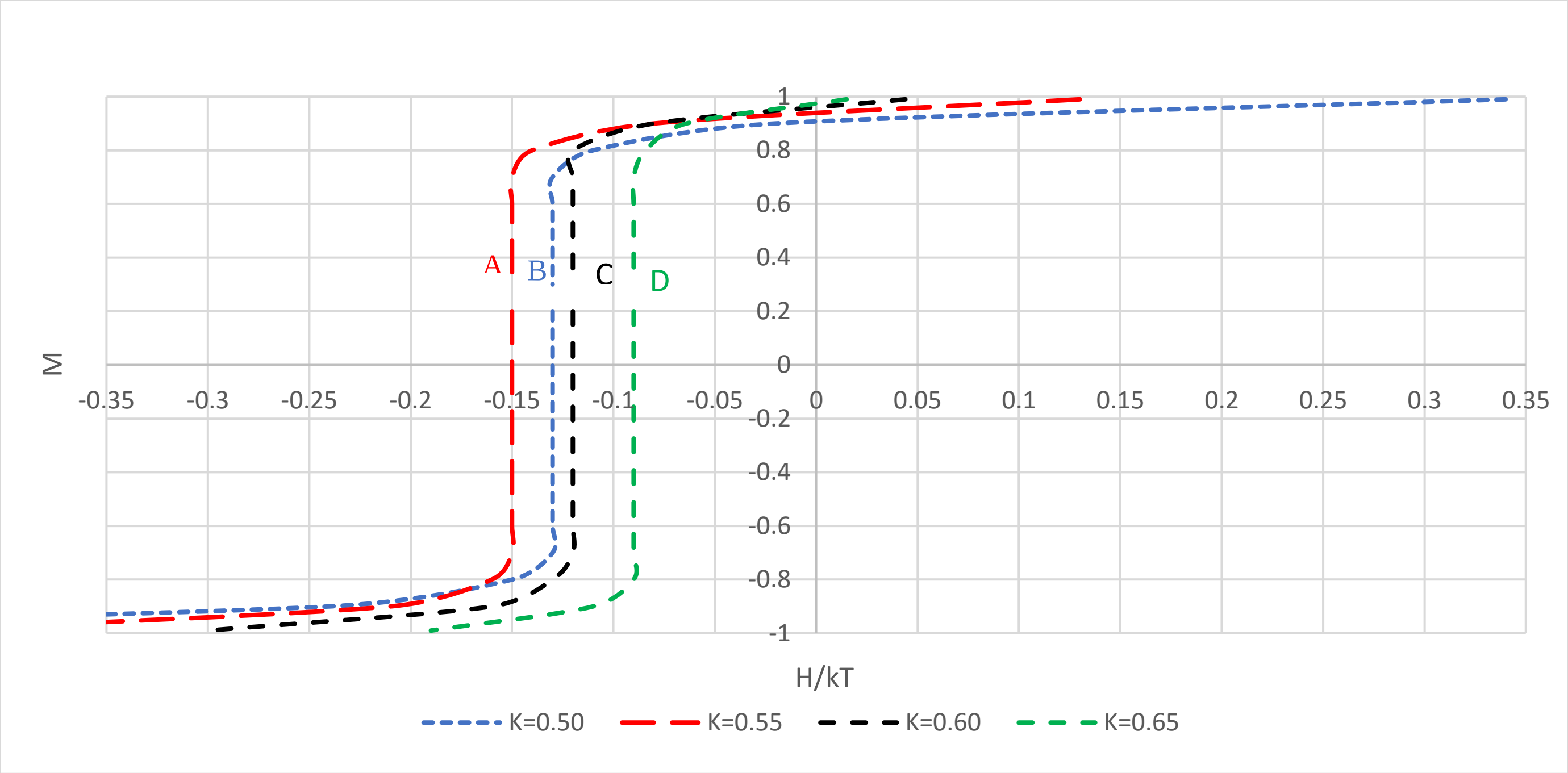

**Figure 4** Dependence of the magnetization on dimensionless magnetic field for various nearest neighbour interaction energies (K=J/kT): (A) 0.50, (B) 0.55, (C) 0.60 and (D) (0.65).

## 5. Summary

By a careful consideration of various arrangement of random distribution of sites, closed form expressions for the number of B-W edges have been deduced for all the 65536 arrangements of a square lattice of 16 sites. This has led to a simplified eqn for the partition function for non-zero values of H and J. By formulating these A(p,q) values in terms of the total number of sites (N), it is now possible to obtain the thermodynamic limit. A semi-empirical eqn for the magnetization has been proposed. The variation of the magnetization with the magnetic field for T> Tc and T < Tc exhibits the expected trend and is in conformity with the experimental data. The exact critical exponents pertaining to the zero-field susceptibility ($\gamma$) and magnetization at the critical temperature ( $\delta$) are also shown to arise from the magnetization eqn.

Acknowledgements

The financial support by MATRICS program of Science and Engineering Research Board, Government of India is gratefully acknowledged.